\def \tr {{\rm tr}}
\documentclass[twocolumn,showpacs,preprintnumbers,amsmath,amssymb]{revtex4}
\usepackage{epsf}
\tighten

\begin{document}
\title{Dynamical shift conditions for the Z4 and BSSN formalisms}
\author{C.~Bona and C.~Palenzuela}
\affiliation{
\\ Departament de Fisica, Universitat de les Illes Balears,
\\     Ctra de Valldemossa km 7.5, 07122 Palma de Mallorca, Spain}

\begin{abstract}
A class of dynamical shift conditions is shown to lead to a
pseudo-hyperbolic evolution system, both in the Z4 and in the BSSN
Numerical Relativity formalisms. This is done by using a
plane-wave analysis which can be viewed as an extension of the
standard Fourier analysis for this kind of systems. The proposed
class generalizes the harmonic shift condition, where light speed
is the only non-trivial characteristic speed, and it is contained
into the multi-parameter family of minimal distortion shift
conditions recently proposed by Lindblom and Scheel. The
relationship with the analogous 'dynamical freezing' shift
conditions used in black hole simulations discussed.
\end{abstract}

\maketitle
\section{Introduction}

General covariance is a characteristic property of Einstein's
theory of Gravitation. There are four coordinate degrees of
freedom in the field equations, allowing to freely choose the
spacetime coordinates $x^{\mu}$ or, in the framework of the 3+1
decomposition, the lapse function $\alpha$ and the shift vector
$\beta^i$.

This gauge freedom can be used, like in the early years of General
Relativity, to set up a complete evolution system (consisting of
the field equations plus the gauge conditions) with a well posed
Cauchy problem. The well known harmonic coordinate conditions
\begin{equation}\label{4harmonic}
    \Box \,x^\mu =
    (g^{\rho \sigma} \nabla_\rho \nabla_\sigma)\, x^\mu= 0\,,
\end{equation}
provide a simple way to get a well posed initial value problem \cite{Choquet55}
with an extremely simple principal part, consisting of one wave equation for every
component of the four-dimensional metric tensor:
\begin{equation}
    \Box \,g_{\mu\nu} = \cdots
\end{equation}
\cite{Fock59, DeDo21}, where the dots stand for terms not
belonging to the principal part.

Although the resulting system is still currently used in
analytical approximations, its use in Numerical Relativity is very
limited, mainly because the four conditions (\ref{4harmonic})
completely exhaust the gauge degrees of freedom, and there is no
flexibility left that could be used to fit the peculiarities of
the specific systems one wants to model (but see also some
generalizations in Refs.~\cite{HE73} to \cite{SW02}).

The current alternative, represented by the 'new hyperbolic
formalisms' (Refs.~\cite{CR83} to \cite{ST02}), is to use somehow
the momentum constraint as a tool for ensuring hyperbolicity,
instead of the three space coordinate conditions in
(\ref{4harmonic}). In this way, only the time gauge condition
\begin{equation}\label{timeharmonic}
    \Box \,x^0 = 0
\end{equation}
is kept, or one of its generalizations (harmonic slicing). This
happens to be very convenient for numerical simulations, because
(\ref{timeharmonic}) implies a direct relationship between the
lapse $\alpha$ and the spatial volume element $\sqrt{\gamma}$,
which can be used to avoid collapse singularities \cite{BM88}. The
main advantage is that one can use now the shift degrees of
freedom to simplify the system, either using normal coordinates
(zero shift) or any other kinematical choice adapted to the
specific problem under study.

Recently, there has been a renewed interest in the use of
dynamical shift vectors. In the context of the BSSN formalism
\cite{SN95, BS99}, some shift conditions have been proposed
\cite{BH02} that manage to 'freeze' black hole dynamics near the
horizon, leading to long term numerical simulations although. As
we will see later, the harmonic shift condition given by
(\ref{4harmonic}) is similar to (but not contained in) the ones
discussed in Ref.~\cite{BH02}.

It is clear that the principal part of the original BSSN system is
modified by the choice of a dynamical shift, although no
hyperbolicity analysis of the modified system has been yet
published, to the best of our knowledge. This is not surprising
because the BSSN formalism, like the ADM one, is of a mixed type:
first order in time, but second order in space, and therefore the
standard Fourier analysis would lead to the conclusion that the
mixed order system is not hyperbolic~\cite{GKO95}. This has been
explicitly shown by Fritelli and Gomez for both the ADM and BSSN
formalisms~\cite{FG00}. We will present here an alternative
plane-wave analysis, based on the underlying physics, in order to
reveal a related property, which we will call
'pseudo-hyperbolicity' to avoid confusion. As far as the
underlying physics does not change when passing from the fully
second order system to the mixed order version of the same
equations, pseudo-hyperbolicity can be seen as the imprint left on
the mixed order system by the true hyperbolicity of the fully
second order version which was at the starting point.

From a different point of view, a generalization of the minimal
distortion shift condition has been proposed by Lindblom and
Scheel~\cite{LS03} in the context of the first order KST formalism
\cite{KST01}. Surprisingly enough, in spite of the fact that the
resulting system contains at least twenty-two free parameters,
none of the cases discussed in \cite{LS03} verifies the condition
that all the non-trivial characteristic speeds do coincide with
light speed. The surprise comes from the fact that one would
expect this condition to be ensured by the use of the full set of
harmonic coordinates (\ref{4harmonic}), which are actually a
particular case of the gauge conditions proposed in \cite{LS03}.

As a contribution to clarify this issues, we will present here a
family of dynamical shift conditions, which is closely related
with the ones presented in \cite{BH02} and \cite{LS03}. We will do
so first in the framework of the general-covariant Z4 formalism
\cite{Z4}, extending then the results to the BSSN case, which can
be derived from the Z4 one in a simple way \cite{BLPZ03}.  We will
perform a complete plane-wave analysis of both systems. The
resulting characteristic speeds are directly related with the main
free parameters of the proposed family. The further requirement
that all the non-trivial characteristic speeds do coincide with
light speed, allows one to recover the harmonic case.

\section{The Z4 System}

\subsection{The Evolution Equations}

The Z4 covariant formalism introduces a four-dimensional vector as
a supplementary dynamical field $Z_\mu$. The evolution equations
are obtained by adding the (symmetrized) covariant derivatives of
$Z_\mu$ to Einstein's field Equations:
\begin{equation}\label{EinsteinZ4}
  R_{\mu \nu} + \nabla_{\mu} Z_{\nu} + \nabla_{\nu} Z_{\mu} =
  8\; \pi\; (T_{\mu \nu} - \frac{1}{2}\, T\; g_{\mu \nu})~.
\end{equation}
In the 3+1 decomposition the line element is written as:
\begin{equation}\label{line_element}
    ds^2 = -\alpha^2\, dt^2 + \gamma_{ij}\, (dx^i + \beta^i\,dt)\,(dx^j + \beta^j\,dt)
\end{equation}
where $\alpha$ and $\beta^i$ are the lapse and the shift,
respectively, and $\gamma_{ij}$ is the spatial 3-metric. Using
this decomposition, the general covariant equations
(\ref{EinsteinZ4}) do consist of a system of pure evolution
equations:

\begin{eqnarray}
\label{Z4dtgamma}
  (\partial_t -{\cal L}_{\beta})~ \gamma_{ij}
  &=& - {2\;\alpha}\;K_{ij}
\\
\label{Z4dtK}
   (\partial_t - {\cal L}_{\beta})~K_{ij} &=& -\nabla_i\alpha_j
    + \alpha\;   [{}^{(3)}\!R_{ij}
    + \nabla_i Z_j+\nabla_j Z_i
\nonumber \\
    &-& 2K^2_{ij}+(\tr K-2\Theta)\;K_{ij}
\nonumber \\
    &-& S_{ij}+\frac{1}{2}\,(\tr\; S -\; \tau)\;\gamma_{ij}\;]
\\
\label{Z4dtTheta} (\partial_t -{\cal L}_{\beta})~\Theta &=&
\frac{\alpha}{2}\; [{}^{(3)}\!R + 2\; \nabla_k Z^k + (\tr K - 2\;
\Theta)\;\tr K
\nonumber \\
&-&  \tr(K^2)  - 2\; Z^k {\alpha}_k/\alpha - 2\tau]
\\
\label{Z4dtZ}
 (\partial_t -{\cal L}_{\beta})~Z_i &=& \alpha\; [\nabla_j\;({K_i}^j
  -{\delta_i}^j \tr K) + \partial_i \Theta
\nonumber \\
  &-&2\; {K_i}^j\; Z_j  -  \Theta\, {{\alpha}_i/ \alpha} - S_i]
\end{eqnarray}
where we have noted
\begin{equation}\label{tauSdef}
  \Theta \equiv  \alpha \; Z^0,~
  \tau \equiv  8 \pi  \alpha^2\; T^{00},~
  S_i \equiv  8 \pi \alpha \; T^0_{\;i},~
  S_{ij} \equiv 8 \pi \;T_{ij}.
\end{equation}

In a recent work \cite{BLPZ03}, a symmetry breaking mechanism is
proposed that, starting from the Z4 system
(\ref{Z4dtgamma}-\ref{Z4dtZ}), allows one to recover an evolution
system which is equivalent, up to quadratic source terms, to the
BSSN system \cite{SN95, BS99} (partial symmetry breaking). Also,
in the first order case, the same mechanism allows one to recover
the multi-parameter KST system \cite{KST01} (full symmetry
breaking) or, to be more precise, a 'live gauge' version of the
same \cite{ST02}.

\subsection{Gauge evolution equations}

The harmonic gauge conditions (\ref{4harmonic}) can be easily
expressed in the 3+1 formalism as
\begin{eqnarray}
  (\partial_t -\beta^{r}\partial_r)~\alpha &=&
  -~\alpha^2\;{\tr} K  \label{harmonicalpha}\\
  (\partial_t -\beta^{r}\partial_r)~\beta^i &=&
  -~\alpha^2\;[\partial^i ln (\alpha\,\sqrt{\gamma})
  + \,\partial_j \gamma^{ij}]\,,   \label{harmonicbeta}
\end{eqnarray}
where the first equation is the (harmonic) slicing condition,
whereas the second one provides the (harmonic) shift once the
slicing is known.

The harmonic slicing condition (\ref{harmonicalpha}) has been
generalized in the context of the Z4 system as follows
\cite{BLPZ03}:
\begin{equation}\label{dtAlpha}
 (\partial_t -\beta^{r}\partial_r)~\alpha = -~\alpha^2\;f\;[{\tr} K
 - m \Theta]\,,
\end{equation}
where the parameter $f$ is directly related with the gauge
propagation speed, whereas $m$ provides a coupling with the
energy-constraint-violating modes, represented by the quantity
$\theta$. We will see in the following analysis that, if one wants
the gauge speed to coincide with light speed ($f=1$), then a
pseudo-hyperbolic system is obtained only if $m=2$, so the
coupling given by the $m$ parameter can not be neglected. This
conclusion coincides with the result of Ref.~\cite{BLPZ03}, where
it was confirmed by the robust stability numerical test.

The harmonic shift evolution equation (\ref{harmonicbeta}) can be
generalized along the same lines:
\begin{equation}\label{dtBeta}
 (\partial_t -\beta^{r}\partial_r)~\beta^i = -~\alpha^2\;[2\,\mu\,V^i
 + a\,\partial^i ln \alpha - d\, \partial^i ln \sqrt{\gamma}]
                       - \eta\, \beta^i
\end{equation}
where we have defined
\begin{equation}\label{V}
  V_i = \partial_i ln \sqrt{\gamma} - \frac{1}{2}
        \partial^j \gamma_{ji} - Z_i
\end{equation}
Notice that the advection term on the left-hand-side, which was
absent in Ref.~\cite{BH02}, is
needed if one wants to recover the harmonic shift as a particular
case. As we will see in the following analysis, the parameters
$\mu$ and $d$ are directly related with the characteristic speeds
of the longitudinal and transverse shift components, respectively,
in the same way as $f$ is related with gauge speed. The parameter
$a$, instead, has no direct relationship with the characteristic
speeds: its role is very similar to the parameter $m$ in the lapse
condition, as we will see that specific values of $a$ will be
required to ensure pseudo-hyperbolicity in degenerate cases, so
one can not just neglect this kind of coupling. The parameter
$\eta$, in turn, corresponds instead to a damping term which has
shown to be crucial to get stable long term simulations
\cite{BH02}. We have not included, however, the $\eta$ term in our
analysis to avoid masking the genuine wave propagation effects
with the artificial damping produced by this kind of terms.

\section{Linear plane-wave analysis}

The system (\ref{Z4dtgamma}-\ref{Z4dtZ}) is of a mixed type: first
order in time but second order in space. This means that,
according to the standard methods~\cite{GKO95}, based on the
Fourier analysis of the principal part, it can not be classified
as hyperbolic. This is also the case of the original ADM
system~\cite{FG00}, where the quantities ($\Theta$, $Z_{k}$) are
supposed to be zero. In what follows, we will present an
alternative plane-wave analysis, starting with the ADM case first
and including the supplementary quantities ($\Theta$, $Z_{k}$)
later.

It is well known that any metric can be written down at a given
spacetime point \textrm{P} in a locally inertial coordinate system
such that the first derivatives of the metric coefficients vanish
at \textrm{P}. We will take advantage of this to write down the
line element at \textrm{P} as
\begin{equation}\label{background}
    ds^2 = -\alpha_{0}^2\, dt^2 + \gamma^0_{\;ij}\,
    (dx^i + \beta_{0}^{\;i}\,dt)\,(dx^j + \beta_{0}^{\;j}\,dt)\,.
\end{equation}
It is clear that the validity of the expression (\ref{background})
is strictly local: second and higher order derivatives of the
metric coefficients at \textrm{P} can not be supposed to vanish:
they are rather related one another by the field equations. This
suggests the splitting of the metric into two components:
\begin{itemize}
    \item A uniform static background of the form (\ref{background})
    \item A dynamical perturbation which, when superimposed to the
    background in a linear way, allows one to recover the full metric.
\end{itemize}

It makes sense then to decompose the dynamical perturbation into
plane waves, with a space dependence given by
\begin{eqnarray}\label{metricdecomp}
  \alpha - \alpha_{0} &=&  e^{i\, \omega \cdot x}\,
  \tilde{\alpha}(\omega, t)  \\
  \beta^k - \beta^{\;k}_0 &=& e^{i\, \omega \cdot x}\,
  {\tilde{\beta}}^k(\omega, t)  \\
  \gamma_{ij} - \gamma_{\;ij}^0 &=& 2\,e^{i\, \omega \cdot x}\,
   {\tilde{\gamma}}_{ij} (\omega, t)
\end{eqnarray}
where $\omega_k = \omega\, n_k$, $\gamma_{\;ij}^0\,n^i\, n^j=1$.

Up to here, we have followed the standard Fourier analysis. Now we
will depart from the standard path by decomposing the dynamical
variable $K_{ij}$ in a form which is consistent with the exact
evolution equation (\ref{Z4dtgamma}), namely
\begin{equation}\label{Kdecomp}
    K_{ij} = i\, \omega\, e^{i\, \omega \cdot x}\,
                   {\tilde{K}}_{ij} (\omega, t)\,,
\end{equation}
where one must notice the $i\,\omega$ factor on the
right-hand-side. This extra factor is not present in the standard
hyperbolicity analysis, where only the principal part of the
system is used, breaking in that way the direct relationship
(\ref{Z4dtgamma}) between the metric and the extrinsic curvature.

Note that equation (\ref{Z4dtgamma}) is crucial to relate the
original (second order in time) version of the field equations
(\ref{EinsteinZ4}) with the resulting (first order in time) 3+1
version (\ref{Z4dtgamma}-\ref{Z4dtZ}). This is why we will choose
the alternative decomposition (\ref{Kdecomp}) in order to look for
the imprint in the 3+1 system  of the hyperbolicity properties of
the original second order version.

The same thing can be done with the supplementary quantities,
which can be considered as an additional perturbation of the
Einstein's equations background. The original system
(\ref{EinsteinZ4}) is of first order in $Z_{\mu}$, but it can be
viewed alternatively as being of second order in some 'potential'
quantities $Y_{\mu}$ which time derivative can be defined to be
precisely $Z_{\mu}$. In this way (\ref{EinsteinZ4}) could be seen
as a fully second order system in ($\gamma_{\mu\nu}$, $Y_{\mu}$)
and the same arguments as before would justify the following
plane-wave decomposition of the supplementary quantities
\begin{eqnarray}
  \Theta &=& i\, \omega\, e^{i\, \omega \cdot x}\,
  \tilde{\Theta} (\omega, t)   \\
  Z_k &=& i\, \omega\, e^{i\, \omega \cdot x}\,
  {\tilde{Z}}_k (\omega, t)\,,
\end{eqnarray}
where the $i\,\omega$ factors still appear on the right-hand-side.

\section{Pseudo-hyperbolicity of the Z4 system with dynamical shift}

We will perform here a linear plane-wave analysis of the Z4
system. This means to substitute the perturbations described in
the preceding section into the evolution equations
(\ref{Z4dtgamma}-\ref{Z4dtZ}, \ref{dtAlpha}, \ref{dtBeta}),
keeping only the linear terms. We get:
\begin{eqnarray}
  (\partial_t -i\, \omega\, \beta^{\;n}_0)~ \tilde{\alpha}
  &=& - i\, \omega\, \alpha_0^2\, f\, [\tr \tilde{K} - m\, \tilde{\Theta}]
\label{Fdtalp} \\
  (\partial_t -i\, \omega\, \beta^{\;n}_0)~ {\tilde{\beta}}^i
  &=& - i\, \omega\, \alpha_0^2\,[(2\, \mu - d)\, n^i\, \tr \tilde{\gamma}
\nonumber \\
      &-& 2\, \mu\, ({\tilde{\gamma}}^{ni} + {\tilde{Z}}^i)
      + a\, n^i\, \tilde{\alpha}/\alpha_0]
\label{Fdtbeta} \\
  (\partial_t -i\, \omega\, \beta^{\;n}_0)~ {\tilde{\gamma}}_{ij}
  &=& - i\, \omega\, [\alpha_0\, {\tilde{K}}_{ij}
  \nonumber \\
  &-& \frac{1}{2}\,(n_i\, {\tilde{\beta}}_j + n_j\, {\tilde{\beta}}_i )]
\label{Fdtgamma} \\
  (\partial_t -i\, \omega\, \beta^{\;n}_0)~ {\tilde{\Theta}}
  &=& - i\, \omega\, \alpha_0\, [\tr \tilde{\gamma} - {\tilde{\gamma}}^{nn}
        - {\tilde{Z}}^n]
\label{FdtTheta} \\
  (\partial_t -i\, \omega\, \beta^{\;n}_0)~ {\tilde{Z}}_i
  &=&  - i\, \omega\, \alpha_0\, [n_i\, (\tr \tilde{K} - \tilde{\Theta})
         - {{\tilde{K}}^n}_i]
\label{FdtZ} \\
  (\partial_t -i\, \omega\, \beta^{\;n}_0)~{\tilde{K}}_{ij}
  &=&  - i\, \omega\, \alpha_0\, [{\tilde{\gamma}}_{ij}
      + n_i\,n_j\,(\tr \tilde{\gamma} + \tilde{\alpha}/\alpha_0)
\nonumber \\
      &-& n_i\, ({\tilde{\gamma}}_{nj} + \tilde{Z}_j)
      - n_j\, ({\tilde{\gamma}}_{ni} + \tilde{Z}_i)]
\label{FdtK}
\end{eqnarray}
where the letter $n$ replacing one index means contracting that
index with $n_i$.

Notice that we have kept the linear source terms in (\ref{Fdtalp},
\ref{Fdtgamma}), in contrast with the usual practice in the
standard Fourier analysis, where only the principal part of the
system is considered. In fact, our plane wave analysis includes
(up to the linear order) all the source terms (with the only
exception of the artificial damping one for the shift, as
discussed before). Therefore, the underlying physics is accounted
for in a consistent way. In particular, the direct relationship
between the metric and the extrinsic curvature is fully preserved.
This means that the characteristic speeds we are going to compute
should be the same ones that could be obtained from either the
fully second order or the fully first order versions, where the
standard Fourier analysis can be applied in a way which is
consistent with the underlying physics of the problem.

The system (\ref{Fdtalp} - \ref{FdtK}) can be written in a compact
way as
\begin{equation}\label{Fus}
    \partial_t \tilde{u} = - i\, \omega\,
             [\mathbf{A} - \beta^{\;n}_0~\mathbf{I}]\,
             \tilde{u}\,,
\end{equation}
where $\tilde{u}$ is the perturbation array. The geometric
properties of matrix on the right-hand-side (characteristic
matrix) are obviously related with the dynamics of the plane-wave
perturbations. Allowing for the trivial structure of the shift
term, the pseudo-hyperbolicity of the evolution system (\ref{Fus})
will depend on the properties of the main matrix $\mathbf{A}$. We
will say that the system (\ref{Fus}) is 'pseudo-hyperbolic' if and
only if $\mathbf{A}$ has real eigenvalues and a complete set of
eigenvectors, that is, the number of independent eigenvectors must
be the same as the number of independent dynamical fields ($20$ in
our case). This is the analogous of the strong hyperbolicity
property of first order systems. The use of the term
'pseudo-hyperbolicity' is just to avoid confusion.

We can start computing the eigenmodes which do not contain shift
terms. In this list we have:
\begin{description}
    \item[Energy cone] There are 2 $\Theta$-related eigenmodes, propagating
    with light speed,
         $v= -\beta_0^{\;n} \pm \alpha_0$\,:
      \begin{equation}\label{Z4EVenergy}
        \tilde{\Theta} \pm (tr \tilde{\gamma}
        - {\tilde{\gamma}}^{nn} - {\tilde{Z}}^n)\,.
      \end{equation}
    \item[Light cones] There are 10 more eigenmodes propagating
    with light speed, $v= -\beta_0^{\;n} \pm \alpha_0$\,:
  \begin{eqnarray}
    \label{Z4EVNTlight} {\tilde{K}}_{na} &\pm& {\tilde{\gamma}}_{na} \\
    \label{Z4EVNNlight} {\tilde{K}}_{ab} &\pm&
    {\tilde{\gamma}}_{ab}\,,
  \end{eqnarray}
    where the letters $a,b$ replacing an index mean the projection
    orthogonal to $n_i$.
    \item[Lapse cone] There are 2 $\alpha$-related eigenmodes
    propagating with
    speed $v= -\beta_0^n \pm \alpha_0\, \sqrt{f}$\,:
      \begin{equation}\label{Z4EVgauge}
        \sqrt{f}~[tr \tilde{K} - B_1\, \tilde{\Theta}] \pm
        [\tilde{\alpha}/\alpha_0 + (2-B_1)\,(tr \tilde{\gamma}
              - {\tilde{\gamma}}^{nn} - {\tilde{Z}}^n)]\,,
      \end{equation}
      where have used the shortcut
    \begin{equation}\label{B1}
    B_1 \equiv \frac{m\,f - 2}{f-1}\,.
    \end{equation}
      The factor $f$ must be greater than zero for
      pseudo-hyperbolicity. Notice that, in the degenerate case
      $f=1$ (harmonic slicing), a well defined pair of eigenmodes
      is obtained only if $m=2$, so that the parameter $B_{1}$ can
      take any value (arbitrary mixing with the energy cone).

\end{description}
The shift-related cones are:
\begin{description}
    \item[Transverse shift cones] There are 4 eigenmodes propagating
    with speeds
         $v= -\beta_0^{\;n} \pm \alpha_0\, \sqrt{\mu}$\,,
      \begin{equation}\label{Z4EVTshift}
        (\tilde{\beta}_a/\alpha_0) \pm
           2\, \sqrt{\mu}~({\tilde{\gamma}}_{na} + \tilde{Z}_a)\,.
      \end{equation}
The factor $\mu$ must be greater than zero for
pseudo-hyperbolicity. Notice that, in the vanishing shift case,
they reduce to the second term, which would correspond to standing
eigenmodes (zero characteristic speed).
\end{description}

\begin{description}
    \item[Longitudinal shift mode] There are 2 eigenmodes propagating with speed
         $v= -\beta_0^n \pm \alpha_0\, \sqrt{d}$\,,
      \begin{eqnarray}\label{Z4EVNshift}
       \sqrt{d} \; &[& \tilde{\beta}^n/\alpha_0 + B_2\, tr \tilde{K}
                     + B_3\, \tilde{\Theta}] \pm
       \nonumber \\
        &[&(a + B_2)\, \tilde{\alpha}/\alpha_0 - d\, tr \tilde{\gamma}
        \nonumber \\
        & &+ (2\, \mu + 2\, B_2 + B_3)\,(tr \tilde{\gamma}
              - {\tilde{\gamma}}^{nn} - {\tilde{Z}}^n) ]
      \end{eqnarray}
We have used again the shortcuts
\begin{equation}\label{B23}
    B_2 \equiv \frac{d - a\, f}{f-d}\,,\;
    B_3 \equiv \frac{(2-m\, f)B_2 +
    2\, \mu - a\, m\,f}{d-1}\,.
\end{equation}
A necessary condition for pseudo-hyperbolicity is that the factor
$d$ should be greater than zero. This condition is also sufficient
in the generic case where $d$ is different from both 1 and $f$.
Notice that in the degenerate cases one would need to impose
additional conditions on the free parameters in order to get a
well defined pair of eigenmodes. For instance, if $d=f$ one must
have $a=1$, so that the parameter $B_2$ can take any value
(arbitrary mixing with the gauge cone). If we have further
degeneracy, that is $d=f=1$ (remember that $f=1$ implies $m=2$),
then it follows from (\ref{B23}) that $\mu=1$ also, so that one
gets the harmonic shift case.
\end{description}

In summary, there are 20 fields in the evolution system and we
have got real characteristic speeds and 20 independent
eigenvectors, provided that all the characteristic speed
parameters $f,\,\mu,\,d$ are greater than zero. The system in then
pseudo-hyperbolic in the generic case, although degenerate cases,
where different characteristic speeds actually coincide, require
additional conditions on the remaining parameters $a,\,m$.

\section{pseudo-hyperbolicity analysis for the BSSN system with dynamical shift}

A pseudo-spectral analysis~\cite{Taylor81} of the original BSSN
system \cite{SN95}~\cite{BS99} has been done recently
\cite{Reu03}. We will proceed here instead with the linear
plane-wave analysis of the complete system, including the
dynamical shift terms. We can take advantage of the symmetry
breaking mechanism proposed in Ref.~\cite{BLPZ03}. Starting from
the Z4 system equations (\ref{Z4dtgamma} to \ref{Z4dtZ}), we will
take the following steps:
\begin{enumerate}
    \item Perform the dynamical fields recombination
       \begin{equation}\label{Ktilde}
           {K'}_{ij} \equiv K_{ij} - \frac{n}{2}\,\Theta\,\gamma_{ij}
       \end{equation}
    \item Suppress $\Theta$ as a dynamical quantity, setting its
    value equal to zero wherever it appears in the evolution
    equations.
\end{enumerate}

This process alters the evolution equation for the extrinsic
curvature $K_{ij}$, even if  one has $K_{ij}={K'}_{ij}\,$ after
the second step. One gets as a result a one parameter family of
evolution systems, with different principal parts for every value
of the $n$ parameter, namely:
\begin{eqnarray}
\label{dtAlpha2}
 (\partial_t -\beta^{r}\partial_r)~\alpha &=& -~\alpha^2\;f\;{\tr}K \\
 \label{dtBeta2}
 (\partial_t -\beta^{r}\partial_r)~\beta^i &=& -~\alpha^2\;[2\,\mu\,V^i
 + a\,\partial^i ln \alpha - d\, \partial^i ln \sqrt{\gamma}]
 \nonumber \\
 &-& \eta\, \beta^i \\
 (\partial_t -\beta^{r}\partial_r)~ \gamma_{ij}
  &=& - {2\;\alpha}\;K_{ij}
  \nonumber \\
  &+& \gamma_{ik}\, (\partial_j \beta^k)
  + \gamma_{jk}\, (\partial_i \beta^k)
\label{Z3dtgamma} \\
 (\partial_t -\beta^{r}\partial_r)~Z_i
  &+& \partial_k [\alpha\, (\delta^k_i\, trK - {K^k}_i )] = \cdots
\label{Z3dtZ}\\
   (\partial_t - \beta^{r}\partial_r)~K_{ij}
   &+& \partial_k [\alpha\, {\lambda^k}_{ij}] = \cdots
\label{Z3dtK}
\end{eqnarray} where we have noted for short
\begin{eqnarray}\label{Z3lambda}
    2\;{\lambda^k}_{ij} = \partial^k \gamma_{ij}
    + \delta^k_i\, (\partial_j \,ln \alpha
          + \partial_j \,ln \sqrt{\gamma} + 2\, V_j)
\nonumber \\
    + \delta^k_j\, (\partial_i \,ln \alpha
          + \partial_i \,ln \sqrt{\gamma} + 2\, V_i)
    - n\, V^k\, \gamma_{ij}
\end{eqnarray}
In particular, it has been shown in \cite{BLPZ03} that the principal
part of the system obtained when $n=4/3$ can be rewritten, by
further rearranging the dynamical fields, as that of the BSSN
system. We will name this system Z3-BSSN to avoid confusion. Both
systems are then linearly equivalent (equivalent up to quadratic
source terms, see Ref.~\cite{Z3}), so that showing
pseudo-hyperbolicity for the Z3-BSSN system, as we will do in the
present section, amounts to show the same for the original BSSN
system.

As far as the second step of the symmetry breaking procedure
suppresses the dynamical field $\Theta$, the linear plane-wave
analysis must be repeated from scratch, although the results of
the preceding section provide a useful guide. We obtain the
following list of eigenvectors and eigenvalues:

\begin{description}
    \item[Standing mode] There is 1 mode propagating along the
    normal lines, that is with speed
         $v= -\beta_0^{\;n}$\,:
      \begin{equation}\label{Z3EVenergy}
      tr \tilde{\gamma} - {\tilde{\gamma}}^{nn} - {\tilde{Z}}^n\,.
      \end{equation}
      This is what is left of the Energy cone in the Z4 system
      after suppressing $\Theta$ as a dynamical field.
    \item[Light cones] There are 10 modes propagating with speed
         $v= -\beta_0^{\;n} \pm \alpha_0$, namely
\begin{eqnarray}
 \label{Z3EVTTlight} {\tilde{K}}_{ab} &\pm&
 [{\tilde{\gamma}}_{ab} - \frac{2}{3}\;
 ( tr \tilde{\gamma} - {\tilde{\gamma}}^{nn} - {\tilde{Z}}^n )\,
 {\tilde{\gamma}}_{ab}^0] \\
 \label{Z3EVNTlight} {\tilde{K}}_{na} &\pm&
 {\tilde{\gamma}}_{na}\,.
\end{eqnarray}
    \item[Lapse cone] There are 2 modes propagating with speed
         $v= -\beta_0^{\;n} \pm \alpha_0\, \sqrt{f}$\,,
      \begin{equation}\label{Z3EVlapse}
        \sqrt{f}\,tr \tilde{K} \pm
        \tilde{\alpha}/\alpha_0
      \end{equation}
    \item[Transverse shift cones] There are 4 modes propagating with speed
         $v= -\beta_0^{\;n} \pm \alpha_0\, \sqrt{\mu}$\,,
      \begin{equation}\label{Z3EVTshift}
        (\tilde{\beta}_a/\alpha_0) \pm
           2\, \sqrt{\mu}\,({\tilde{\gamma}}_{na} + \tilde{Z}_a)
      \end{equation}
    \item[Longitudinal shift cone] There are 2 modes propagating with speed
         $v= -\beta_0^n \pm \alpha_0\, \sqrt{d}$:
      \begin{eqnarray}\label{Z3EVNshift}
        & &\sqrt{d} \;[\tilde{\beta}^n/\alpha_0 + B_2\, tr \tilde{K}] \pm
        \nonumber \\
        & &[(a + B_2)\, \tilde{\alpha}/\alpha_0 - d\, tr \tilde{\gamma}
          + 2\, \mu \,(tr \tilde{\gamma}
              - {\tilde{\gamma}}^{nn} - {\tilde{Z}}^n) ]\;
      \end{eqnarray}
\end{description}
where the parameter $B_2$ is the same that appears in the Z4
system, as defined in (\ref{B23}). Notice that, in the degenerate
case $d=f$, a well defined pair of eigenmodes is obtained only if
$a=1$, so that $B_2$ can take any value (arbitrary mixing with the
lapse cone).

In summary, there are 19 fields in the evolution system and we
have got real characteristic speeds and 19 independent
eigenvectors, provided that all the characteristic speed
parameters $f,\,\mu,\,d$ are greater than zero. The system is then
pseudo-hyperbolic in the generic case, although the degenerate
case $d=f$ requires the additional condition $a=1$. Notice that
the harmonic case is recovered precisely when
\begin{equation}\label{harmoniccase}
    d=f=\mu=1\,.
\end{equation}

Comparing with Ref.~\cite{BH02}, the only relevant term in their
"Gamma driver" shift conditions is the $\tilde{\Gamma}^i$ one. It
is clear that this corresponds to our parameter choices
\begin{equation}\label{gammadrivershift}
     a=0\;,\;\;\;\;\;\mu = \frac{3}{4}\,d\,
\end{equation}
so that pseudo-hyperbolicity is ensured provided that $d \neq f$.
The main difference, as stated before, is that our shift
conditions (\ref{dtBeta}) are a generalization of the harmonic
ones (\ref{harmonicbeta}). This is only possible by keeping, as we
do, the advection term in the shift evolution equation
(\ref{dtBeta}). This term was suppressed in Ref.~\cite{BH02} in
order to freeze the dynamics in the neighborhood of the black
hole's apparent horizon.

\section{Conclusions}

In this paper, we have studied a multi-parameter family of
dynamical gauge conditions (\ref{dtAlpha}, \ref{dtBeta}), which
generalizes the harmonic gauge conditions (\ref{4harmonic}) along
the ways sketched in Refs.~\cite{BH02},~\cite{LS03} (Gamma-driver
shift conditions). Starting with (the second order version of) the
general covariant Z4 formalism, we have computed explicitly all
the eigenmodes, identifying the choices of the gauge parameters
$\{f,m,\mu,a,d\}$ that make the full evolution system
pseudo-hyperbolic. The relationship between the gauge parameters
and the characteristic speeds is direct and simple. Depending on
the parameters choice, gauge (lapse and shift) propagation speeds
can be made to coincide or not with light speed.

The same kind of analysis has been done with the corresponding
gauge conditions, for the second order system (\ref{dtAlpha2} -
\ref{Z3dtK}), which is linearly equivalent to that of the BSSN
system~\cite{Z3}. The Gamma-driver shift condition which has been
used in Ref.~\cite{BH02} corresponds to the parameters choice
(\ref{gammadrivershift}), but with the advection term in the
left-hand-side of the shift equation (\ref{dtBeta2}) suppressed.
One could redo the analysis without that term, just by adding it
to the right-hand-side of the same equation. This would affect the
shift cones (\ref{Z3EVTshift}, \ref{Z3EVNshift}), which would then
intersect the other ones, leading to a more involved causal
structure. We have chosen to keep instead the shift advection term
in place, so that the original harmonic gauge condition
(\ref{4harmonic}) is kept inside our family.

Finally, we will briefly discuss the relevance of our results in
connection with the ones presented in Ref.~\cite{LS03}, in the
context of the (first order) KST formalism. In that paper, a
22-parameter family of gauge conditions was presented which
contains (\ref{dtAlpha2}, \ref{dtBeta2}) as a sub-family, but no
parameter choice was found ensuring both symmetric hyperbolicity
and the additional requirement that all the non-trivial
characteristic speeds should coincide with light speed. Is this
because these two conditions are actually incompatible in the KST
formalism... or it is rather because there are only few 'good'
choices (maybe just one), hidden in the huge 22-parameter space,
just waiting to be identified?.

We can not fully answer this question here because we are dealing
in this paper just with pseudo-hyperbolicity, not with the
stronger condition of symmetric hyperbolicity for first order
systems. However, there is a direct relationship between the Z3
systems proposed in Ref.~\cite{Z3} and the KST formalism. This
means that we can at least provide necessary conditions that can
be helpful by reducing parameter space in the quest for a
definitive answer:
\begin{itemize}
    \item The harmonic case is the only one in which we can get
    both pseudo-hyperbolicity and light speed as the only
    nontrivial characteristic speed. We can easily
    identify the harmonic case in the family of gauge conditions
    proposed in Ref.~\cite{LS03}, getting the following
    restrictions on their gauge parameters:
    \begin{equation}\label{gaugeharm}
    \mu_S=2\,,\; \mu_L=2\sigma=1\,,\;
    \epsilon_S=-1\,,\; \epsilon_L=0\,,\; \lambda=-1\,.
    \end{equation}
    As far as the damping terms containing
    $\{\kappa_S\,,\;\kappa_L\}$ are not relevant for the
    hyperbolicity analysis, this completely fixes the gauge
    parameters.
    \item As seen in the preceding sections, the Light
    cones can be always obtained independently of the details of
    the lapse and shift cones. This means that the results presented
    in Ref.~\cite{Z3} for the zero shift case, where the
    non-trivial characteristic speeds were given explicitly in terms
    of the KST parameters, can be applied as such. It follows that
    the light speed condition imposes the following additional
    restrictions on the KST parameters:
    \begin{equation}\label{KSTharm}
    \eta = 4\,(1+\chi)\,,\;\;\;\;\;\chi = 2 \gamma\,(1+\chi).
    \end{equation}
    This means that $\zeta$ and $\gamma$ (which amounts to the
    parameter $n$ in (\ref{Z3dtK}): $n=-2\gamma$) are the only remaining
    (independent) KST parameters , although $\zeta=-1$ is usually
    required for symmetric hyperbolicity~\cite{Z4}.
\end{itemize}
There are, of course, the 10 additional coupling parameters $\psi$
introduced in Ref.~\cite{LS03}, so that the question can not be
completely solved here. Nevertheless, we hope that the conditions
we provide, which actually reduce by half the 22 parameter space,
will pave the way to a definitive answer.

{\em Acknowledgements: This work has been supported by the EU
Programme 'Improving the Human Research Potential and the
Socio-Economic Knowledge Base' (Research Training Network Contract
HPRN-CT-2000-00137), by the Spanish Ministerio de Ciencia y
Tecnologia through the research grant number BFM2001-0988 and by a
grant from the Conselleria d'Innovacio i Energia of the Govern de
les Illes Balears. We acknowledge Denis Pollney for the valuable
comments and discussions during his visit to Palma.}

\bibliographystyle{prsty}

\end{document}